\documentclass[epj]{webofc}
\usepackage[varg]{txfonts}   
\usepackage{hyperref}

\renewcommand*{\eqref}[1]{Eq.~(\ref{eq:#1})}

\newcommand*{\figref}[1]{Fig.~(\ref{fig:#1})}
\newcommand*{\figlab}[1]{\label{fig:#1}}

\newcommand*{\seclab}[1]{\label{sec:#1}}
\wocname{\includegraphics[width=0.25cm,clip]{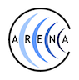} ARENA2018}
\wocname{ARENA2018}
\woctitle{ARENA2018}
\woctitle{\includegraphics[width=0.25cm,clip]{logoARENA} ARENA2018}
\begin{document}
\title{Radio Morphing - towards a fast computation of the radio signal from air-showers}
%
%
%
\author{\firstname{Anne} \lastname{Zilles}\inst{1}\fnsep\thanks{\email{zilles@iap.fr}}
\and \firstname{Olivier} \lastname{Martineau-Huynh}\inst{2}
\and \firstname{Kumiko} \lastname{Kotera}\inst{1}
\and \firstname{Matias} \lastname{Tueros}\inst{3}
\and \firstname{Krijn} \lastname{de Vries}\inst{4}
\and \firstname{Washington} \lastname{Carvalho Jr.}\inst{5}
\and \firstname{Valentin} \lastname{Niess}\inst{6}}
\institute{Sorbonne Universit\'{e}, UPMC Univ.  Paris 6 et CNRS, UMR 7095, Institut d'Astrophysique de Paris, 98 bis bd Arago, 75014 Paris, France 
\and
           LPNHE, CNRS-IN2P3 et Universit\'{e}s Paris VI \& VII, 4 place Jussieu, 75252 Paris, France 
\and
           Instituto de F\'{i}sica La Plata - CONICET/CCT- La Plata. Calle 49 esq 115. La Plata, Buenos Aires, Argentina
\and
           Vrije Universiteit Brussel, Physics Department, Pleinlaan 2, 1050 Brussels, Belgium 
\and
           Universidade de Santiago de Compostela, 15782 Santiago de Compostela, Spain 
\and
           Clermont Universit\'{e}, Universit\'{e} Blaise Pascal, CNRS/IN2P3, Laboratoire de Physique Corpusculaire, BP. 10448, 63000 Clermond-Ferrand, France 
          }

\abstract{%
  Over the last decades, radio detection of air showers has been established as a
promising detection technique for ultrahigh-energy cosmic rays and neutrinos. Very large or
dense antenna arrays are necessary to be proficient at collecting information about these particles and understanding their properties accurately. The exploitation of such arrays requires to run massive air-shower simulations
to evaluate the radio signal at each antenna position, taking into account features such as
the ground topology. In order to reduce computational costs, we have developed a fast computation 
of the emitted radio signal on the basis of generic shower simulations, called
Radio Morphing. The method consists in the calculation of the radio signal of any air-shower by
i) a scaling of the electric-field amplitude of a reference air shower to the target shower, ii)
an isometry on the simulated positions and iii) an interpolation of the radio pulse at the
desired position. This technique enables one to gain many orders of magnitude in CPU time
compared to a standard computation. In this contribution, we present this novel tool and explain its
methodology. In particular, Radio Morphing will be a key
element for the simulation chain of the Giant Radio Array for Neutrino Detection (GRAND)
project, that aims at detecting ultra-high-energy neutrinos with an array of 200 000 radio
antennas in mountainous regions.
}
\maketitle
\section{Introduction}
\label{intro}
\vspace{-0.2cm}
The calculation of the expected radio signal from an extensive air-shower can be performed by macroscopic or a microscopic approaches. The first one describes the radio emission mainly by modeling and fitting of the electric current and charge distribution in a shower. Therefore, this mostly analytical calculation of the signal, as used in e.g. MGMR~\cite{Scholten08} or EVA~\cite{EVA}, is very fast. On the other hand, some parameters of the models which are determining the signal have to be tuned, affecting the prediction of the electric fields. 
The microscopic approach, as used in CoREAS~\cite{CoREAS} or ZHAireS~\cite{Zhaires}, considers each single electron and positron in an air shower individually. Here, derived from first principle, no assumptions on the emission mechanisms are made and there are no free parameters to tune. Due to the underlying Monte-Carlo simulation of the air shower, this approach is highly time-consuming, especially for lower thinning levels, so that the limitations of computational resources are quickly reached.
Especially at early stages of very large radio arrays, e.g. in the case of performance studies of particular topographies for the GRAND experiment~\cite{GRAND}, a more portable and at the same time fast method is needed.

Therefore, we developed \textit{Radio Morphing} which allows to obtain the signal expected by any air shower at any desired observer position, based on the simulation of one generic shower and morphing according to the desired shower parameters via purely mathematical operations.

\section{Method and comparison to microscopic simulations}
\seclab{sec-1}
\vspace{-0.2cm}
The basic idea of Radio Morphing is that the electric field vector $\bf E$$_B($$\bf x'$$ , t')$ of a the radio signal emitted by any shower $B$ and measured at a given position $\bf{x'}$ and time $t'$ can be derived from that of a reference shower $A$ by a set of simple mathematical operations that are applied on the electric field vector $\bf E$$_A($$\bf x$$, t)$ and on its corresponding position $\bf{x}$.\\
It can be found that at fixed distances from the shower maximum $X_{\rm max}$, the strength of the radio signal solely depends on the primary energy ${\cal E}$, the shower direction described by the zenith $\theta$ and the azimuth $\phi$ as well as on the injection height $h$ of the air shower.
Therefore, we simulate the radio signal of a generic shower for antenna positions at fixed distances from $X_{\rm max}$, arranged in a so-called star-shape pattern, with observer positions distributed along the $(\bf{v}\times \bf{B})$- and $(\bf{v}\times \bf{v}\times \bf{B})$-axis, with $\bf{v}$ as the direction of the shower and $\bf{B}$ as the orientation of the Earth's magnetic field.
Following expressions of the electric field amplitude derived by \cite{Allan:1971} and the recent measurements performed by AERA and LOFAR~\cite{AERAEnergy,LOFARMass}, the electric field amplitude can be expressed as a product of factors $j$ scaling linearly with respective shower parameters, most notably the primary energy ${\cal E}$ $(j_{\cal E})$ and the sine of the geomagnetic angle $\alpha$ $(j_\alpha)$.
The zenith $\theta$ and the injection height $h$ of the shower affect the signal strength via the dependency of the air density and the refractive index on the actual height of  $X_{\rm max}$.
From fits to simulations, we found the following density scaling ($j_\rho$):
$|\vec{E}|^2 \propto [\rho_{X_{\rm max}}(h,\theta)]^{-1}$, with $\rho_{X_{\rm max}}(h,\theta)$ as the air density at $X_{\rm max}$.
The density at $X_{\rm max}$ also impacts the value of the refractive index $n$, affecting the size of the Cherenkov ring.
For a decreasing refractive index with height, the area illuminated by the radio signal at a fixed distance from $X_{\rm max}$ decreases while the total emitted energy is conserved. To account for that, we use a factor $j_{\rm C} = 1/k_{\rm C}\ $ with $k_{\rm C}$ expressing the relative value by which the Cherenkov ring increases ($k_{\rm C}>1$) or shrinks ($k_{\rm C}<1$). The simulated antenna positions of the reference shower are adjusted by $k_{\rm C}$.
At a fixed distance to the shower maximum $X_{\rm max}$, the signal of a reference and a target shower can be linked by the showers parameters, leading to the following dependency of the electric field vector:
\begin{center}
$\bf E$$_B($$\bf x'$$, t) = J_{AB}({{\cal E},\theta,\phi,h}) \,$$\bf E$$_A[k_{AB}({\theta, h})\,$$\bf x$$, t]$.
\end{center}
In the scaling, independent effects related to the primary energy, the geomagnetic field, the air density, and air refraction index are taken into account as a scaling matrix $J_{AB}({{\cal E},\theta,\phi,h})=j_{\cal E} j_\alpha j_\rho j_c $ and a factor $k_{AB}({\theta, h})=k_{\rm C}$.
After the scaling of the electric-field amplitude, the scaled observer positions $\bf x'$ of the reference shower have to be adjusted by an isometry to the geometry of the target shower. Profiting from the reference simulations of the radio signal at observer positions arranged in the star-shape pattern, this is done by a simple rotation of the position accordingly to the target shower direction and their translation accordingly to the shower maximum $X_{\rm max}$ of the target shower. 
To receive the signal at the desired antenna positions, we perform a linear interpolation of amplitude and phase of the signal between the simulated observer positions surrounding the desired one in the frequency domain as described in~\cite{EwaThesis}.
%
\begin{figure}[tb]
\centering
\includegraphics[width=12cm,clip]{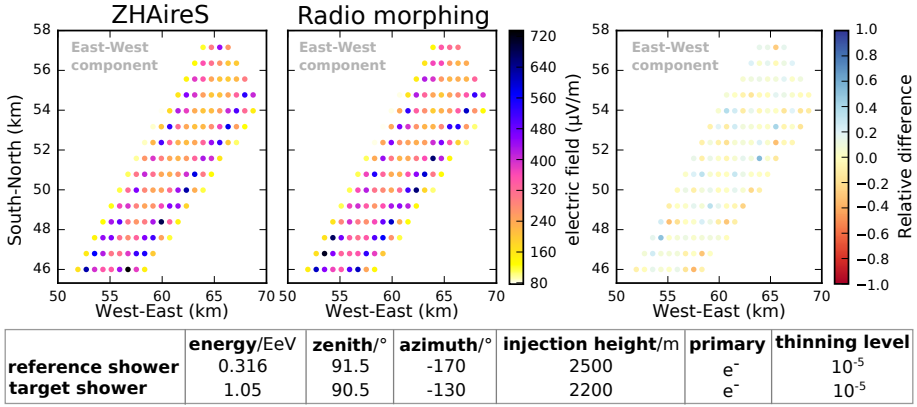}
\caption{Top: Comparison of a ZHAireS simulation (left) and Radio Morphing (center). Right: Relative difference between the peak-to-peak amplitudes at each observer position defined as $(E_{\mbox{zhaires}} - E_{\mbox{rm}}) /E_{\mbox{zhaires}}$.
Bottom: Table of primary's parameters for the reference and target shower. }
\figlab{fig-1}      
\end{figure}
A comparison between the output from ZHAireS and Radio Morphing in the frequency band of $0-500\,\mbox{MHz}$ 
is displayed in~\figref{fig-1} (top).
As target and reference we have simulated air showers with the parameters as given in~\figref{fig-1} (bottom) and a thinning level of $10^{-5}$.  For comparison, the ZHAireS simulation took $\sim 6\,$h for $206\,$observer positions, while via Radio Morphing, the expected radio signal for this target shower was calculated for exactly the same observer positions within $\sim 1\,$min on the same computer. 
The comparison of the peak-to-peak amplitude distribution shows that expected features in the footprint as the Cherenkov ring get reproduced. 
Even though all shower parameters of the reference shower had to be scaled to the target parameters, we found a difference of only $\sim 10$\% in the signal strength for most of the simulated locations, while saving a factor of $\mathcal{O}(2)$ in time.
The maximal differences in the calculated peak amplitudes for single antenna positions  of $\sim 20$\%  can be found for positions near the edges of the Cherenkov cone. Due to the exponential drop of the signal strength at the edges of the cone, already a slight misalignment in the exact position of the cone lead to this order of difference.\\
\\
\noindent
\small{\textbf{Acknowledgement}\\
\noindent
This work is supported by the APACHE grant (ANR-16-CE31-0001) of the French Agence Nationale de la Recherche and by the grant \#2015/15735-1 of the São Paulo Research Foundation (FAPESP).
}
\vspace{-0.35cm}

\end{document}